\documentclass[twocolumn,pre,showpacs]{revtex4}
\usepackage[pdfpagemode=None,pdfstartview=FitH]{hyperref}
\usepackage{graphicx}
\usepackage{amsmath}
\usepackage{epstopdf}

\begin{document}

\def\br{{\bf r}}
\def\vcmi{v_{cm,i}}

\def\Deltar{\Delta  {\bf r}}
\def\Deltav{\Delta {\bf v}}
\def\L{{\bf L}}
\def\r{{\bf r}}
\def\v{{\bf v}}
\def\f{{\bf f}}
\def\p{{\bf p}}
\def\k{{\bf k}}

\title{Internal dissipation of a polymer}

\author{J. M. Deutsch}
\affiliation{
Department of Physics, University of California, Santa Cruz, California 95064}

\date{\today}

\begin{abstract}
The dynamics of flexible polymer molecules are often assumed to be
governed by hydrodynamics of the solvent. However there is considerable
evidence that internal dissipation of a polymer contributes as well. Here
we investigate the dynamics of a single chain in the absence of solvent
to characterize the nature of this internal friction. We model the
chains as freely hinged but with localized bond angles and
3-fold symmetric dihedral angles.  We show that the damping is close
but not identical to Kelvin damping, which depends on the first temporal and
second spatial derivative of monomer position. With no internal potential between monomers,
the magnitude of the damping is small for long wavelengths and weakly
damped oscillatory time dependent behavior is seen for a large range
of spatial modes. When the size of the internal potential is increased,
such oscillations persist, but the damping becomes larger. However
underdamped motion is present even with quite strong dihedral barriers
for long enough wavelengths.
\end{abstract}

\pacs{
}

\maketitle

\section{Introduction}

Polymer molecules in solution are usually modeled assuming that damping is mediated
by interaction with surrounding fluid, or other polymer chains. However it has been
noted early on, that there is another source of dissipation, 
internally generated by the nonlinear dynamics of the polymer chain~\cite{DeGennesBook,Cerf,CerfExp}. In a set
of important experiments~\cite{CerfExp} extensional
relaxation for different solvent viscosities was measured. By extrapolating
the solvent viscosities to zero, an interesting residual dissipation was 
discovered that is internal in origin.  A number of explanations for it have been
proposed~\cite{MacInnes,degennesfriction}, but the origins of this effect
are still not well understood.

In this work, the internal friction of a polymer is examined by considering it
in isolation, that is, without any solvent. This not only sheds light on
the case of polymer solutions, but also on more recent experimental 
techniques, discussed below, used to characterize polymers by first gasifying them without
damaging their integrity.

Developments in mass spectrometry of long chain polymers, have
become important in recent years in order to characterize large biological
molecules such as proteins~\cite{Hillenkamp}. The process involved in these experiments
puts single chains into a vacuum.  The study of polymer damping in this
environment would elucidate the understanding of internal dissipation but
has yet to be studied experimentally.  However recent work has studied
the behavior of single chain molecules in a vacuum theoretically and
by means of computer simulation~\cite{DeutschVacPRL,DeutschExactVac}.
The statistical properties of single chains with no damping, and
subject only Newton's laws was investigated. Both ideal chains (where
only chain connectivity is included and no intra-chain interactions),
and self-avoiding chains, were considered.  In related work, the
detailed chaotic properties of small chains with rigid links has also
been observed for self-interacting chains with two body Lennard-Jones
potentials~\cite{Taylor}. Through a comprehensive analysis of this problem, they
were able to show that the simulated chains were in excellent agreement with
exact predictions from a microcanonical average, providing strong
evidence that the dynamics are indeed ergodic. The chaotic motion derives from both the bond
constraints and the convex nature of scattering. In another related work~\cite{mossa}
computer simulations for polymer molecules were performed to
examine Lyapunov exponents. These were shown to be very small for systems close
to a phase transition, such as the coil-globule transition.

An ideal chain modeled with only linear springs connecting adjacent
monomers is integrable and shows no damping. Energy in every mode
is conserved and cannot be exchanged with other modes. Dissipation
results from the nonlinear coupling of modes to each other. This is
closely related to the problem of nonlinear one dimensional chains
that have been studied extensively using the Fermi, Pasta, and Ulam
(FPU) model~\cite{FPU}. In this model, there is energy transfer 
between modes, however depending on initial conditions, the time
it takes to lose correlation with its initial state can be very
long~\cite{BermanIzrailev}. An ideal chain is a similar one dimensional
system where the displacement of each particle is, in this case, much larger and is vectorial
in nature. However we expect that the same equilibration problems persist.
Therefore to thermalize a polymer in a vacuum efficiently,
the model chosen~\cite{DeutschVacPRL} uses
rigid links so that the local motion would be highly chaotic. It also mimics
the fact that the fluctuations in bond lengths are small compared to those involved with
rotational degrees of freedom. On large
length scales however, one might expect universal behavior, independent
of the form of the potential connecting adjacent monomers. As will
be seen below, this appears to be the case.

The dynamics of such ideal chains are also closely related to the study
of the dynamics and energy flow in one dimensional chains, that govern
its anomalous conductivity~\cite{NarayanRamaswamy,DeutschNarayan}.
The dynamics in one dimension systems with momentum conservation show faster
relaxation than in higher dimensions due to Galilean invariance.
A consequence of momentum conservation in higher dimensions
is responsible for long time tails in liquids that were predicted
theoretically~\cite{AlderWainwright} and confirmed experimentally
\cite{Carneiro}.  In the case of one dimensional nonlinear chains, a wide
range of different models give the same heat conductivity exponent, for example
the Sinai Pencase model~\cite{pencase}, the Random Collision
Model~\cite{DeutschNarayan} and FPU chains~\cite{MaiDharNarayan}.
The main difference between these one dimensional systems and ideal chains in a vacuum 
is that the former has a fractal dimension of 1, and the latter, of 2.
It is therefore hard to define local hydrodynamic variables as a function
of position for polymer chains. However
conservation of energy and momentum are common to both problems. 

In addition, in the case of an ideal chain in a vacuum,
the equilibrium properties are effected by the conservation of angular momentum~\cite{DeutschExactVac}, and 
this can be analyzed exactly, showing that when the angular momentum is zero, the radius of gyration is significantly
smaller than without that conservation law enforced. 

Initial results~\cite{DeutschVacPRL} for an ideal chain in a vacuum showed that 
its dynamics are very different than those of a chain in solution.
A freely hinged chain of $N$ links, with constant link lengths was studied.
The time auto-correlation function for position oscillates
and is slowly damped. The damping time appears to scale as
$T_{rel} \propto N^{(1.85\pm.15)}$, where $L$ is the chain length. 

The reason for this behavior can be seen by understanding
the form of dissipation such chains should have. Because of Galilean
invariance, the frictional force $f_d$ on a monomer cannot be proportional
to its velocity, as this would imply that a uniformly translating
chain would slow down. If we denote the position of a chain at
arclength $s$ as ${\bf r}(s)$, then the simplest term respecting
this translational invariance is 
\begin{equation}
\label{eq:KelvinFriction}
f_d \propto \frac{\partial^3 {\bf r}}{\partial t\partial^2 s}
\end{equation}
This is the form of ``Kelvin
Damping"~\cite{SethnaBookKelvinFriction}. This theoretical form was
proposed by MacInnes based on a calculation done by him for a model
system~\cite{MacInnes}. It was later  proposed as the origin of
Cerf friction~\cite{Cerf} where it was argued that this form was
compatible with experiments characterizing internal
friction~\cite{CerfExp}.  In Fourier space
with wave-vector $k$ conjugate to $s$, and $\omega$ conjugate to $t$,
this dissipation is proportional to $i\omega k^2 {\hat r}(k,w)$. Longer
wavelength modes are weakly damped, implying underdamped motion for
long enough wavelengths as we will shortly see. Using this damping along with linear forces
between monomers, analogous to the Rouse model~\cite{Rouse}, one
has
\begin{equation}
\rho \frac{\partial^2 {\bf r}}{\partial t^2} =
\Big(\kappa + C \frac{\partial}{\partial t} \Big) \frac{\partial^2 {\bf r}}{\partial s^2}  +
{\bf \xi(s,t)}
\label{eq:rouselike}
\end{equation}
Here $\xi$ is a noise term added to maintain ideal chain statistics. 
$\rho$ is the mass per unit arclength.  
The term $\kappa r''(s)$ is identical to that of the Rouse equation
and describes the net force experienced from neighboring segments. For
Gaussian chains, $\kappa = 3 k_B T$, where  $T$ is the temperature~\cite{DeGennesBook}.
For a linear chain, $r'(s) = 0$ at the two ends and the chain can be
expanded in terms of Fourier modes, similar to the analysis of the Rouse
equation
\begin{equation}
{\bf r}(s) =  \sum_k {\hat {\bf r}_k} \cos(k s)
\end{equation}
with $k = n \pi/L$,  $n = 1,2,\dots$.
so that
\begin{equation}
\label{eq:model}
\rho \frac{\partial^2 {\bf \hat r}}{\partial t^2} =
-k^2 \Big(\kappa - C \frac{\partial}{\partial t} \Big) {\bf \hat r}  + {\bf \hat \xi}(k,t)
\end{equation}
When noise is omitted, this is of the form a damped harmonic
oscillator with an effective mass of $\rho$, a spring constant of $ K = \kappa k^2$ and
friction of $\nu = Ck^2$, so that
\begin{equation}
\label{eq:DampedHO}
M \ddot{r} + \nu\dot{r} + K r = 0
\end{equation}
For underdamped long wavelength modes, 
the solution is proportional to $\exp(i\Omega t)$  with 
$\Omega_k \equiv \omega_k + i\lambda_k$. The
correlation function for a $k$ mode can then be calculated

\begin{equation}
\label{eq:corrdef}
\langle r_k(0)r_k^*(t)\rangle  = \langle|r_k|^2\rangle {\rm Re}~(\exp(i\Omega_k t)\Omega_k/\omega_k) .
\end{equation}

$\lambda_k \propto k^2$ (in the underdamped regime) so the damping time $\propto 1/k^2$. The frequency of 
oscillation, $\omega_k$, is proportional to $k$ for small $k$ . Therefore in this
limit, there are many oscillations, $O(k^{-1})$ in a damping time.

The model of a freely hinged chain is not realistic at a microscopic level. A
molecule such as polyethylene has a strong orientational dependence as dihedral
angles are varied. The chain will spend most of its time in minima and make
transitions between these. The arguments above are general and do not give
any indication of the prefactor for the dissipation term. This is expected
to depend on the details of the potential. We will use numerical methods
to see to what extent Eq. \ref{eq:model} is satisfied and
to determine how the size of the prefactor $C$ depends on the potential used.
We will find, surprisingly, that underdamped motion is still present
even with rather strong potentials.  We first describe the numerical method
used in this work and then give the simulation results in the following section.

\section{Numerical Method}
\label{sec:numericalmethod}

This method is an extension of the simulation of rigid link systems developed
by the author previously~\cite{DeutschScience,DeutschMadden} which considered
the case of a highly overdamped systems where inertia was negligible. Here
we consider the general case of particles with mass $m$ and damping $\gamma$.
In the subsequent sections, we will take $\gamma = 0$.

The coordinates are denoted $\r_1,\r_2,\dots,\r_N$, where $N$ is the
number of masses in the system. The time derivative of these,
that is the velocities, are denoted $\v_1,\v_2,\dots,\v_N$.
As in the earlier work, we will assume that each mass is being acted
on by a force $\f_i$, which can be due to self-interaction or externally
applied. 
It will be convenient to define the difference operator of adjacent coordinates or velocities
$\Delta_i \r \equiv  \r_{i+1} - \r_i$ and $\Delta_i \v \equiv  \v_{i+1} - \v_i$. To keep the
link lengths constant, we
Introducing Lagrange multipliers $t_1,\dots,t_{N-1}$ which describe the tensions between
neighboring masses, we can write the equations of motion as:
\begin{eqnarray}
\label{eq:f=ma}
m \dot{\v_i} + \gamma \v_i & = & t_i (\Delta_i \r) - t_{i-1}(\Delta_{i-1}\r)  + \f_i\\
\dot{\r_i} & = & \v_i
\end{eqnarray}
for $1 < i < N$.
For a linear chain, one can define $t_0 = t_N = 0$ to give the equations for
the chain ends, that is for $i=1$ and $i=N$. We will discuss ring chains below.

We wish to evolve the $\r$'s and $\v$'s in time by finding the tension at every time
step which allows us to iterate the above equations by some appropriate integration scheme. 
However this can be problematic due to the cumulation of errors, as
we require that the magnitude of $\Delta_i \r$ remain very close to the step length, $l$.
Define the error in this quantity as
\begin{equation}
\epsilon_i \equiv |\Delta_i \r|^2 - l^2 .
\end{equation}
Without the problem of numerical error, the tensions would be determined
by the condition $\epsilon_i = 0$ for $i=1,\dots,N$. By differentiating
this equation twice with respect to time, a formula for the tensions can
be obtained. However numerical error will cause bond lengths to stray from
their initial values. Therefore we need to introduce feedback into the method so
that non-zero $\epsilon_i$ will be pushed back towards zero. So instead we consider the equation
\begin{equation}
\label{eq:feedback}
A\epsilon_i + B\dot{\epsilon_i} + \ddot{\epsilon_i} = 0
\end{equation}
where $A$ and $B$ are constants causing a damping of errors with time, 
and whose values are determined to maximize computational efficiency.
This can be rewritten as
\begin{equation}
\label{eq:feedback_rewritten}
2\Delta_i \r\cdot\Delta_i\dot{\v} = -A\epsilon_i - 2B\Delta_i \r\cdot \Delta_i \v - 2|\Delta_i \v|^2 
\end{equation}

Applying the difference operator $\Delta_i$ to Eq. \ref{eq:f=ma} and taking the dot
product with $\Delta_i \r$ gives
\begin{align}
 m    \Delta_i  \r &\cdot \Delta_i \dot\v  +  \gamma \Delta_i \r \cdot \Delta_i \v 
\nonumber \\
 =  &t_{i+1} \Delta_i \r \cdot \Delta_{i+1} \r +
 t_{i-1} \Delta_i \r \cdot \Delta_{i-1} \r -
\nonumber \\
 & 2  t_i |\Delta_i \r|^2 + \Delta_i \r \cdot \Delta_i \f 
\end{align}

Putting this into Eq.  \ref{eq:feedback_rewritten} gives
\begin{align}
\label{eq:tridag}
t_{i+1} \Delta_i \r&\cdot\Delta_{i+1} \r   -  2 t_i |\Delta_i \r|^2  +  t_{i-1} \Delta_i \r \cdot \Delta_{i-1} \r
\nonumber \\
   = & \frac{m}{2}   ( 
 -A\epsilon_i - 2B\Delta_i \r\cdot \Delta_i \v - 2|\Delta_i \v|^2) 
\nonumber \\
  & -    \Delta_i \r \cdot \Delta_i \f + \gamma \Delta_i \r \cdot \Delta_i \v
\end{align}

The left hand side contain the tensions, which are unknowns, but the rest
of the variables and the right hand side are all known. These equations
form a tridiagonal matrix equation which can be solved for the $t_i$'s
in a time $O(N)$.

After the tensions are determined, they are used in the right hand side of
Eq. \ref{eq:f=ma}. These $2N$ first order differential equation
can be integrated by a variety of methods. This paper uses fourth
order Runge Kutta to do this. Excluding the operations involving the computation
of the forces $\f$, this algorithm runs in $O(N)$
operations per time step. 
For short range potentials, such as are used
here, the forces also require $O(N)$ operations, meaning the operations
per time step are $O(N)$. This scales the same way with $N$ as non-constrained simulations such
as molecular dynamics with variable bond lengths. 

In general, this algorithm can be easily extended
to systems with any set of connections between masses, such as branched topology. 
For the simplest variant, the ring chain, we modify the problem by introducing
fictitious particles $\r_{N+1} \equiv \r_1$, and $\r_{0} \equiv \r_N$.
This places an additional link between monomer $1$ and $N$, giving rise to
an additional tension $t_N$. To keep the form of the equations the same,
it is convenient to introduce $t_0 \equiv t_N$.
In this case Eq. \ref{eq:tridag} becomes a cyclic tridiagonal
system of equations that can be easily solved as well. However for more
general connections, the order of the number of operations will be greater than $N$.

Typically, the step size used in this work for a Runge Kutta iteration was $.01$. Therefore this
algorithm provides an efficient method for investigating dynamics of a polymer
in a vacuum.

\section{Simulations}
\label{sec:simulations}

\begin{figure}[htp]
\begin{center}
\includegraphics[width=\hsize]{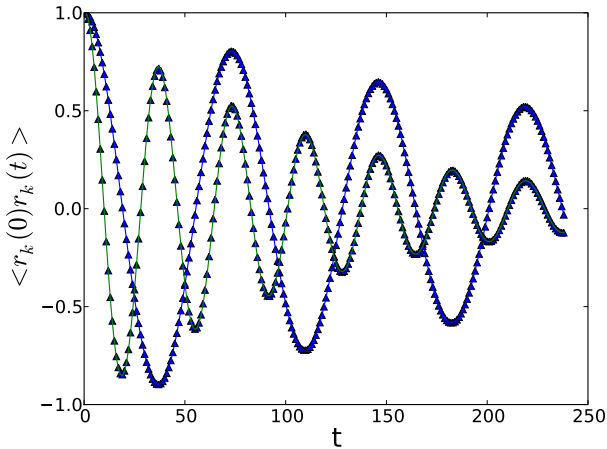}
\caption{ (Color Online) 
Correlation functions for a freely hinged chain in a vacuum for $N=64$ for
the lowest two modes, shown by the solid triangles. The line going through them are fits using
Eq. \ref{eq:corrdef}. 
}
\label{fig:CorrFitIdeal}
\end{center}
\end{figure}

We first consider the case where there are no potentials but only the freely
hinged constraint. In this work, we consider a step length $l = 1$ and at temperature
$k_B T = 1$. We fit the correlation functions for numerical data for $N=64$ using
Eq. \ref{eq:corrdef}. The first two mode are shown in Fig. \ref{fig:CorrFitIdeal}
As can be seen, the fit to the data
is excellent. This was averaged over $37578$ runs, so that the error
bars on the data are negligible. Clearly for this range of parameters,
the decay of a single $k$-mode is well described by a damped harmonic
oscillator. All data for correlation functions shown are normalized
to unity at $t=0$. This is because the amplitude of the modes is proportional
to $1/k^2$, making it hard to discern the data without this rescaling.

It is important to note that in general, this kind of correlation function
can not be described by a single mode, and that the answer is expected to
be the sum over a large number of modes. In fact, for strong enough dihedral
potentials more modes need to be included, however we shall see that for 
quite strong potentials, a single mode fits the simulation data quite well.

\begin{figure}[htp]
\begin{center}
\includegraphics[width=\hsize]{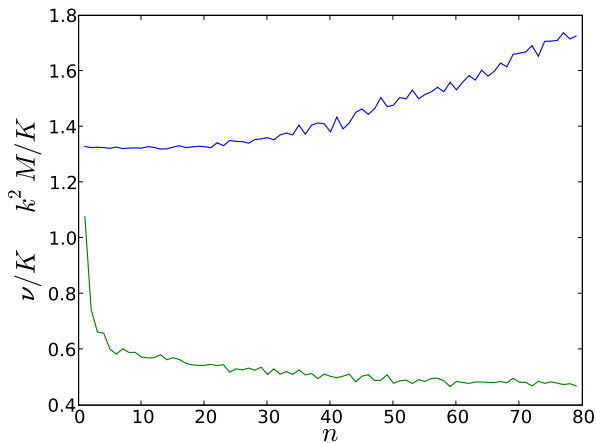}
\caption{ (Color Online) 
The parameters for fits of the time auto-correlation function, Eq. \ref{eq:corrdef}. 
to a damped harmonic oscillator are shown for a freely hinged
chain with $N=256$. The upper line represents the mass $k^2 M/K$ of each mode, and
the lower curve represents that damping $\nu/K$.
}
\label{fig:IdealParams}
\end{center}
\end{figure}

The parameters for the oscillator can be fit as a function of $k$. 
$\Omega_k$ is first determined and then the corresponding parameters
in Eq. \ref{eq:DampedHO}, $M/K$ and $\nu/K$ are calculated.
In Fig. \ref{fig:IdealParams}, $M$ and $\nu$ were fitted for 
different mode numbers $k = \pi n/N$ with $N = 256$ (the step length has been set to unity). 
The fluctuations give a measure of the uncertainty in this data as was checked by fitting with
independent data.

Note that according to Eq. \ref{eq:rouselike}, both curves should be constant as 
a function of the mode number. However a clear variation is seen in each. The effective
mass of the mode increases with increasing $k$, and the effective friction coefficient
is not $\propto k^2$, but is slightly less. The data could be fit equally well, with
a slightly smaller exponent so that $\nu \propto k^p$ with $p < 2$, or as a logarithmic correction, $k^2/\log(k)$.
In either case, the assumption of a local term for the friction is apparently not
completely correct. There is a subtle transfer of energy between modes leading to
non-locality. The same is apparently true for the mass, and no doubt these
two effects are related. The goal of the present work is not to explain these effects
theoretically, but to investigate how the friction coefficient is altered by local potentials
along the chain. Therefore we will now turn our attention to this problem.

\begin{figure}[htp]
\begin{center}
\includegraphics[width=\hsize]{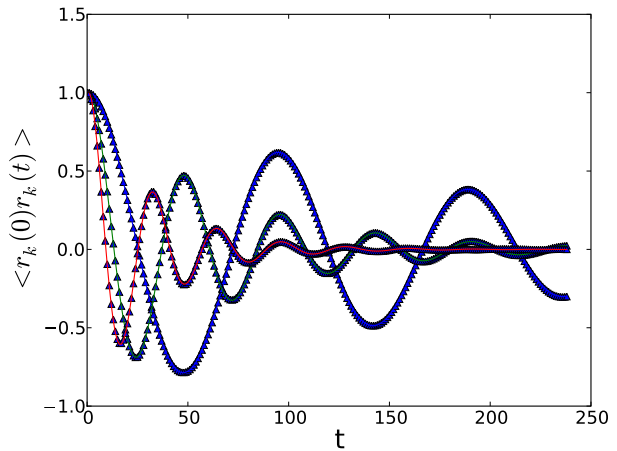}
\caption{ (Color Online) 
Time auto-correlation functions where bond angle is now restricted for $N=64$.
Data for the first three modes is shown by the solid triangles. The lines going through it is a fit using
Eq. \ref{eq:corrdef}. 
}
\label{fig:CorrFitAzim0}
\end{center}
\end{figure}

\begin{figure}[htp]
\begin{center}
\includegraphics[width=\hsize]{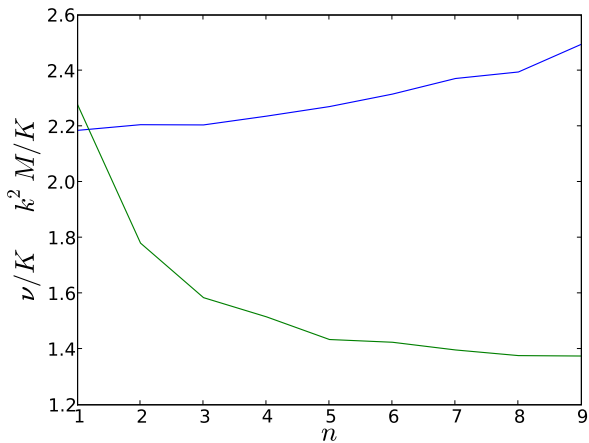}
\caption{ (Color Online) 
The parameters for fits of the time auto-correlation function, Eq. \ref{eq:corrdef},
to a damped harmonic oscillator are shown for the same data as in Fig. \ref{fig:CorrFitAzim0}.
The upper curve is a measure of the mass $k^2 M/K$, and the lower curve is measure of the damping
$\nu/K$.
}
\label{fig:Azim0Params}
\end{center}
\end{figure}

The oscillations seen are a result of the weak form of damping at low
wave-number. Do these persist if the chain is no longer freely hinged? To answer
this, we first examine a model where the distribution of bond, or valency angles, is 
weighted around a set of values centered at a particular angle $\theta_V$.
To restrict configurations in this manner, a potential is constructed between
the nearest neighbors of monomer $i$ as follows:
\begin{equation}
U_{val} = \frac{U_b}{4} (|{\bf r}_{i+1}-{\bf r}_{i-1}|^2 - r_0^2)^2 .
\end{equation}
Because the distance between nearest neighbors is fixed, 
for large $U_b$, this will limit configurations
as just described. $r_0$ is chosen to give the value of $\theta_V$
desired, in this case $\theta_V = 104\,^{\circ}$.
This value is somewhat arbitrary and varies according to the chemical structure of the
polymer. 
As a result, the dynamics will be restricted
as well. Fig. \ref{fig:CorrFitAzim0} shows the time auto-correlation
function for $U_b = 10$, for the first three modes.
In this case we see again that the fit to Eq. \ref{eq:corrdef} is excellent. The long
wavelength modes are still quite underdamped. A fit of the harmonic oscillator parameters
is shown in Fig. \ref{fig:Azim0Params}. It is similar to the freely hinged case,
Fig. \ref{fig:CorrFitIdeal}, suggesting that the variation of the drag coefficient
with k may be understandable by some general mechanism.

\begin{figure}[htp]
\begin{center}
\includegraphics[width=\hsize]{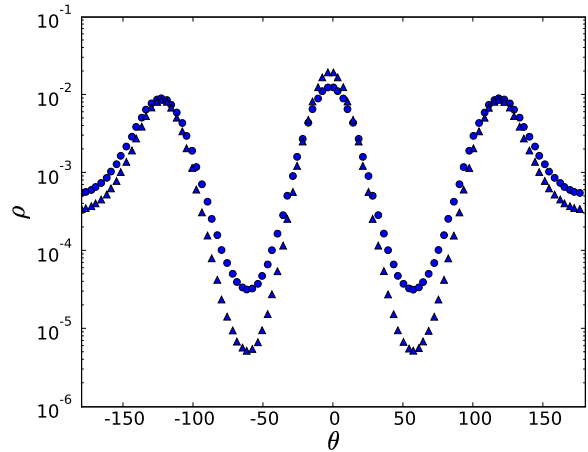}
\caption{ (Color Online) 
The distribution of dihedral bond angles for two different heights of the
   dihedral potential, $V_d = 3$ (circles) and $V_d = 4$ (triangles). Here
   $U_b = 10$ and $N=64$.
}
\label{fig:ThetaDist}
\end{center}
\end{figure}

\begin{figure}[htp]
\begin{center}
\includegraphics[width=\hsize]{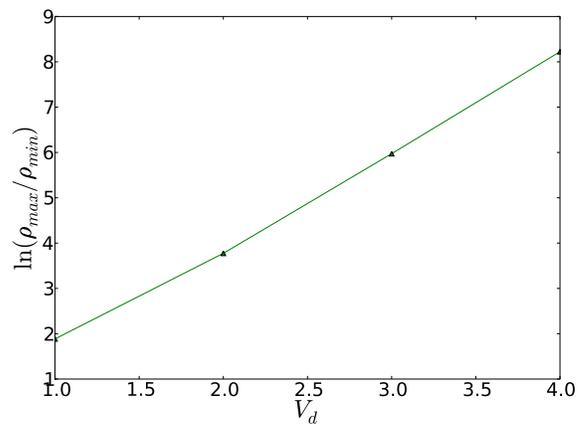}
\caption{ (Color Online) 
The natural logarithm of the
dihedral angle distribution maximum to minimum as a function of potential amplitude $V_d$ with $U_b = 10$.
}
\label{fig:RhoVsVd}
\end{center}
\end{figure}

Usually as a link is rotated, there are three energy minima as a function
of the dihedral angle.  To make this model more realistic, it is necessary
to add such a dihedral potential. As a function of the dihedral angle
$\theta$, between two adjacent monomers, the potential energy $U =
V_d \cos(3\theta)$. This can be expressed in terms of $\cos\theta$
using the usual trigonometric identity. If the bond angle is fixed,
$\cos\theta$ can be expressed in terms of dot products which is
computationally advantageous. Because the bond angle is almost constant,
we still use the same dot product formula in the simulation. This also
breaks symmetry between the three minima moving the trans state slightly
relative to the two gauche states. This is also seen in real data such as
for polyethylene~\cite{PolymerCharacterization}. Fig.  \ref{fig:ThetaDist}
shows the distribution of bond angles, $\rho(\theta)$, for two separate dihedral potentials,
with $V_d = 3$ and $V_d = 4$, for $U_b = 10$ and $N=64$.
As the amplitude of the dihedral potential, $V_d$ increases, the ratio
of the maximum for the distribution, $\rho_{max}$ to the minimum
$\rho_{min}$ increases. In these simulations the energy scale was chosen so that the temperature is $1$
and Fig.  \ref{fig:RhoVsVd} shows that $\log(\rho_{max}/\rho_{min}) \approx 2 V_d$,
It is reduced from approximately $4 V_d$ because of coupling to
other degrees of freedom such as bond angle.
For a real temperature of $330 K$, comparison with earlier
modeling~\cite{BoydGeeHanJin} of polyethylene, yields that $V_d \approx
2.5$ (as we are using units where $T=1$.) At lower real temperatures,
the potential will be correspondingly higher. 

\begin{figure}[htp]
\begin{center}
\includegraphics[width=\hsize]{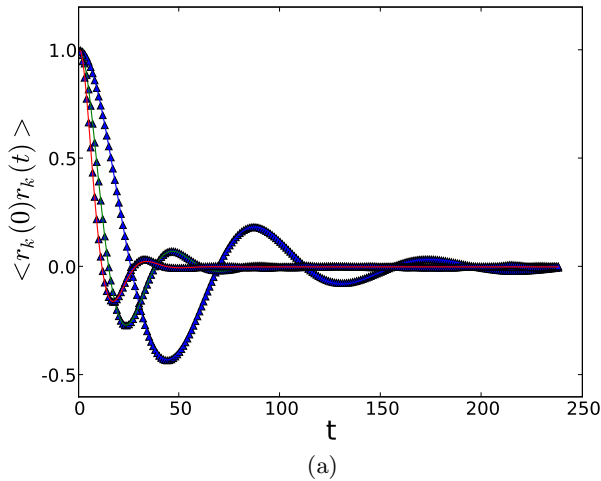} \\
(a) \\
\includegraphics[width=\hsize]{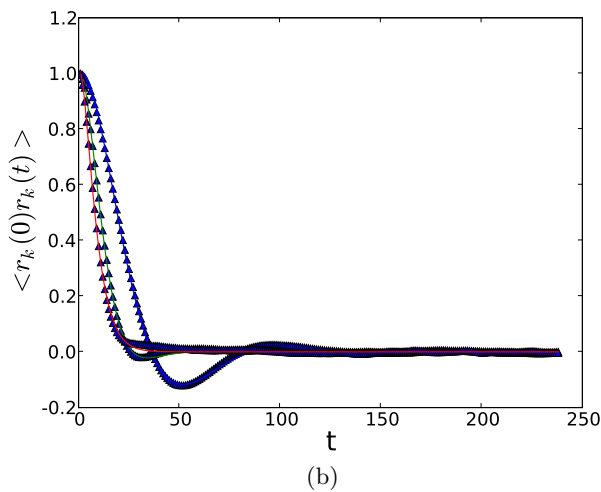} \\
(b)
\caption{ (Color Online) 
Time auto-correlation functions where bond angle and a dihedral potential for $N = 64$, $U_b = 10$. 
Data for the first three modes is shown by the solid triangles. The lines going through it is a fit using
Eq. \ref{eq:corrdef}. (a) with $V_d = 2$  and (b) with $V_d = 3$.
}
\label{fig:CorrFitAzim2and3}
\end{center}
\end{figure}

Fig. \ref{fig:CorrFitAzim2and3} shows the correlation function fit to
the damped harmonic oscillator for $V_d = 2$ (a), and $V_d = 3$ (b). Here
$N=64$ and $U_b = 10$.
In both cases, clear oscillations can be seen for the lowest modes.

\begin{figure}[htp]
\begin{center}
\includegraphics[width=\hsize]{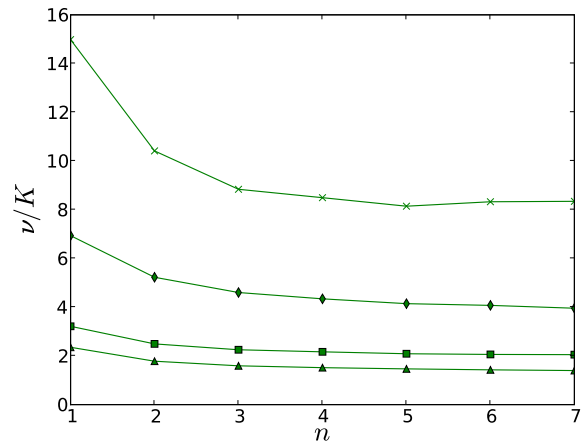}
\caption{ (Color Online) 
The damping as a function of mode number for a range of dihedral potentials with $N=64$.
The bottom curve shown by the triangles is $V_d =0$, the next highest, $V_d =1$ (squares), then $V_d = 2$ (diamonds),
and $V_d = 3$ (crosses).
}
\label{fig:l64Damping}
\end{center}
\end{figure}

It is of interest to know how the frequency dependent damping depends on the height of the
dihedral potential. Fig. \ref{fig:l64Damping} shows the results of fitting different modes
to the damped harmonic oscillator, taking into account the possibility of both underdamped and overdamped
motion. As with Fig. \ref{fig:IdealParams}, the damping $\nu$
is divided by $K \propto k^2$. The non-constant nature of the results show that for
quite substantial $V_d$, there is a similar deviation from the expected Kelvin damping
as was seen in Figs. \ref{fig:IdealParams} and \ref{fig:Azim0Params}.
Note that even if higher modes are overdamped we expect the for lower enough
$k$, and long enough chains, the motion will become underdamped. This is because of the 
$k$-dependent form of the damping $\nu$, which goes to zero for as $k \rightarrow 0$. 

As can be seen from Fig. \ref{fig:RhoVsVd}, for $V_d > 1$, barriers
to rotation are substantial. It would then first appear that motion
would be much more like that of a lattice model, where a monomer
stays in a potential minimum for a long time and flips to another
state. This would suggest that the motion would be heavily overdamped,
with correlation functions like those of the Rouse model. However we
have seen that even in these cases, the motion for long wavelengths,
is underdamped. For strong enough potentials such as  $V_d =4$, the
motion even at long wavelengths and $N=64$, ceases to be underdamped. The
relaxation becomes very slow and does not fit well to a simple harmonic
oscillator.  Many more frequency modes must be considered in this case.
However as we argued above, for long enough chain length, we expect to see
underdamped motion for low $k$.  When the value of $U_b$ is increased,
this increases $\log(\rho_{max}/\rho_{min})$ and will make the barrier
height larger. Still for $N = 64$, when the dihedral constant $U_b$
is raised to $U_b = 40$, oscillations persist when $V_d = 2$.

\section{Conclusions}
\label{sec:conclusions}

These results are of interest for two reasons. First, it sheds
like on the nature of internal friction of polymer chains in solution.
This work characterizes the nature of this  dissipation. It is quite
similar to that of Kelvin friction, Eq. \ref{eq:KelvinFriction}.
However it is not identical as the damping is larger for small
wavenumber. This departure is at present unexplained and is probably
quite nontrivial to understand as it is related to the dynamics
of nonlinear one dimensional chains~\cite{FPU,BermanIzrailev}.

Second, it is important in the understanding the internal dynamics
of polymers in a vacuum of which there is at present little
experimental or theoretical understanding. In reality polymers in this
situation will be charged and have van der Waals interactions. An
investigation of these have shown that their effects are very
important~\cite{DeutschVacPRL}. However as with the understanding of
ideal chains in polymer solutions, it is important to understand an
ideal chain in a vacuum as the starting point for further analysis.

The author thanks Onuttom Narayan for useful discussions.

\end{document}